# London penetration depth in the tight binding approximation: Orthorhombic distortion and oxygen isotope effects in cuprates.

M V Eremin, I A Larionov and I E Lyubin

Department of Physics, Kazan State University, Kazan 420008, Russia

E-mail: Larionov.MRSLab@mail.ru

We present a simple derivation of an expression for the superfluid density $n_s \propto 1/\lambda^2$ in superconductors with the tight binding energy dispersion. The derived expression is discussed in detail because of its distinction from the known expressions for ordinary superconductors with parabolic energy dispersion. We apply this expression for the experimental data analysis of the isotope effect in London penetration depth parameter $\lambda$ in the BiSrCuO and YBaCuO family compounds near optimal doping, taking into account the orthorhombic distortion of crystal structure, and estimate the isotopic change of hopping parameters from the experimental data. We point out that $1/\lambda^2$ temperature behaviour is very sensitive to the ratio $2\Delta_m(T=0)/k_B T_c$ and estimate this quantity for a number of compounds.

PACS: 74.70.-b, 74.72.-h

## 1. Introduction

The basics of superconductor electrodynamics is given by the London equation, $\mathbf{j} = -\dfrac{c}{4\pi\lambda^2}\mathbf{A}$, which describes the relation between the superconducting current density $\mathbf{j}$ and the vector potential $\mathbf{A}$. The parameter $\lambda$ is typically measured through the effective magnetic field penetration depth in a superconductor and gives important information about the microscopic properties. The elaborated microscopic theory for the superfluid density ($n_s \propto 1/\lambda^2$) for ordinary low temperature superconductors is described in [1,2]. The situation for new superconductors is not yet settled. Up to now, different expressions have been employed in order to describe $1/\lambda^2$ data in copper oxide high temperature superconductors (HTSC's) (see, for example, refs. [3-16]). These curcumstances lead to confusion and misunderstanding in interpretation of the temperature dependencies of superfluid density. In the present report, in order to make the situation as clear as possible in HTSC, we perform a simple derivation of the expression for $1/\lambda^2$ in the tight binding approximation, which is widely accepted on the basis of the Angle Resolved Photoemission Electron Spectroscopy (ARPES) data [17].

## 2. Current operator

It is known [18] that the charge transfer amplitude from point $R_l$ to point $R_j$ is proportional to

$$\exp\left(-\mathrm{i}\frac{e}{\hbar c}\int_{R_l}^{R_j}\mathbf{A}\,\mathrm{d}\mathbf{s}\right) \approx \exp\left(-\mathrm{i}\frac{e}{\hbar c}A_x R_{jl}^x\right). \tag{1}$$

Here it is assumed that the field is applied along the x-axis. Any transfer integral in the direction $n_x$ gains the factor

$$t_{jl}^{x} \Rightarrow t_{jl} \exp\left(-i\frac{e}{c\hbar}A_x R_{jl}^{x}\right) \cong t_{jl}\left[1 - i\frac{e}{c\hbar}A_x R_{jl}^{x} - \frac{1}{2}\left(\frac{e}{c\hbar}A_x R_{jl}^{x}\right)^2 + ...\right]. \qquad (2)$$

We consider first the linear correction for the kinetic energy operator of the system:

$$\delta H_{kin}^{(1)} = -i\frac{e}{c\hbar}\sum_{n,l,\sigma} t_{nl} A_x R_{nl}^{x} a_{n,\sigma}^{+} a_{l,\sigma}, \qquad (3)$$

Here $a_{n,\sigma}^{+}$ ($a_{l,\sigma}$) is the creation (annihilation) operator of a quasiparticle at site $n$ ($l$) and $\sigma = \pm 1/2$ – spin quantum numbers.

By comparison of the expression (3) with the energy in the field of the vector-potential,

$$\delta H_{kin}^{(1)} = -\frac{1}{c}\sum_{\mathbf{q}} j_x(-\mathbf{q}) A_{\mathbf{q}}^{x} + h.c., \qquad (4)$$

we may obtain a general expression for the Fourier component of the current density operator $j_x(-\mathbf{q})$. Substituting the mean value of the vector-potential in the harmonic expansion form, following [1],

$$A_x(R_{nl}^{0}) = \frac{1}{2} A_{\mathbf{q}}^{x}(e^{i\mathbf{q}\cdot\mathbf{R}_n} + e^{i\mathbf{q}\cdot\mathbf{R}_l}) + h.c., \qquad (5)$$

In (3), performing a Fourier transformation, $a_{l,\sigma} = \frac{1}{\sqrt{N}}\sum_{\mathbf{k}} a_{\mathbf{k},\sigma} \exp(i\mathbf{k}\cdot\mathbf{R}_l)$, and comparing then with (4), we obtain,

$$j(\mathbf{q}) = -\frac{e}{2\hbar}\sum_{\mathbf{k},\sigma}\left[\frac{d\varepsilon_{\mathbf{k}}}{dk_x} + \frac{d\varepsilon_{\mathbf{k+q}}}{d(k_x+q_x)}\right] a_{\mathbf{k},\sigma}^{+} a_{\mathbf{k+q},\sigma}. \qquad (6)$$

Here $\varepsilon_{\mathbf{k}} = \sum_i t_{ij} \exp(i\mathbf{k}\cdot\mathbf{R}_{ji})$ is the usual expression for the quasiparticle energy in the tight binding approximation, which, after performing the summation over lattice, we take in the form suggested in Ref. [19].

$$\begin{aligned}\varepsilon_{\mathbf{k}} &= \tfrac{1}{2}t_1[(1+\delta_t)\cos k_x a + (1-\delta_t)\cos k_y b] + t_2 \cos k_x a \cos k_y b \\ &+ \tfrac{1}{2}t_3[(1+\delta_t)\cos 2k_x a + (1-\delta_t)\cos 2k_y b] \\ &+ \tfrac{1}{2}t_4(\cos 2k_x a \cos k_y b + \cos 2k_y b \cos k_x a) + t_5 \cos 2k_x a \cos 2k_y b\end{aligned}, \qquad (7)$$

where $t_1$, $t_2$, $t_3$, $t_4$ and $t_5$ – are the effective hole hopping parameters in the $CuO_2$ layer and the parameter $\delta_t$ accounts for the orthorhombic distortion of crystal structure.

Note that in the parabolic zone approximation, $t_{\mathbf{k}} = \varepsilon_{\mathbf{k}} = (\hbar\mathbf{k})^2/2m$, where $m$ is the effective carrier mass, the equation (6) has the form given in a standard textbooks [1],

$$j_x(\mathbf{q}) = -\frac{\hbar e}{2m}\sum_{\mathbf{k},\sigma}[2k_x + q_x] a_{\mathbf{k},\sigma}^{+} a_{\mathbf{k+q},\sigma}. \qquad (8)$$

Hence, one may consider the expression for the current operator (6) as a natural generalization of the well-known expression (8). The latter is valid only in either the weak coupling approximation or in the case of a parabolic zone with an isotropic effective mass of charge carriers.

**3. Mean value of the paramagnetic current**

According to the hands-on terminology in the theory of superconductivity, equation (6) corresponds to the paramagnetic current. The diamagnetic current component is due to the vector-potential quadratic corrections to hoppings (see equation (2)) and will be considered in section 4. In the first approach the mean value of the operator (6) over the ground state is equal to zero. The equation for the London magnetic field penetration depth $\lambda$ in superconductor can be obtained by taking the mean value of equation (6) right up to the second perturbation term. One of the possible calculation schemes is to take the unperturbed ground state wave functions and to use the linear response theory and the Green function technique. The other way is to take into account the changes in the superconductor's ground state due to the external field and then take an average in the linear vector-potential limit. We will use the second scheme because it does not require the Green function technique and because of its simplicity.

Adding $\delta H_{\text{kin}}^{(1)}$ to the Hamiltonian of a superconductor with the linear vector-potential terms we have

$$H = \sum_{\mathbf{k},\sigma} E_{\mathbf{k}} \alpha_{\mathbf{k},\sigma}^+ \alpha_{\mathbf{k},\sigma} - i\frac{e}{c\hbar} \sum_{n,l,\sigma} t_{nl} A_{\mathbf{q}}^x e^{-i\mathbf{q}\mathbf{R}_n} R_{nl}^x a_{n,\sigma}^+ a_{l,\sigma} + h.c.$$

$$= \sum_{\mathbf{k},\sigma} E_{\mathbf{k}} \alpha_{\mathbf{k},\sigma}^+ \alpha_{\mathbf{k},\sigma} + \frac{eA_{\mathbf{q}}^x}{2c\hbar} \sum_{\mathbf{k},\sigma} \left[ \frac{d\varepsilon_{\mathbf{k}+\mathbf{q}}}{d(k_x+q_x)} + \frac{d\varepsilon_{\mathbf{k}}}{dk_x} \right] a_{\mathbf{k},\sigma}^+ a_{\mathbf{k}+\mathbf{q},\sigma} + \frac{e(A_{\mathbf{q}}^x)^*}{2c\hbar} \sum_{\mathbf{k},\sigma} \left[ \frac{d\varepsilon_{\mathbf{k}+\mathbf{q}}}{d(k_x+q_x)} + \frac{d\varepsilon_k}{dk_x} \right] a_{\mathbf{k}+\mathbf{q},\sigma}^+ a_{\mathbf{k},\sigma},$$  (9)

Here $\alpha_{\mathbf{k},\sigma}^+ (\alpha_{\mathbf{k},\sigma})$ are Bogoliubov's creation (annihilation) quasiparticle operators [1,2], $E_{\mathbf{k}} = \sqrt{(\varepsilon_{\mathbf{k}} - \mu)^2 + |\Delta_{\mathbf{k}}|^2}$ is the quasipartical energy, $\Delta_{\mathbf{k}}$ is the complex superconducting gap parameter, and $\mu$ is the chemical potential. The quantities $R_{nl}^x$ have been incorporated in derivatives $\frac{d\varepsilon_{\mathbf{k}+\mathbf{q}}}{d(k_x+q_x)}$ and $\frac{d\varepsilon_{\mathbf{k}}}{dk_x}$. The correction terms can also be expressed through the Bogoliubov's operators. Since the expressions in the square brackets in (9) are odd functions with respect to the transformation $\mathbf{k} \rightarrow -\mathbf{k}$, it is convenient to consider the difference:

$$a_{\mathbf{k},\uparrow}^+ a_{\mathbf{p},\uparrow} - a_{-\mathbf{p},\downarrow}^+ a_{-\mathbf{k},\downarrow} = \left(u_{\mathbf{k}} u_{\mathbf{p}} + v_{\mathbf{k}} v_{\mathbf{p}}\right) \left(\alpha_{\mathbf{k},\uparrow}^+ \alpha_{\mathbf{p},\uparrow} - \alpha_{-\mathbf{p},\downarrow}^+ \alpha_{-\mathbf{k},\downarrow}\right) - \left(u_{\mathbf{k}} v_{\mathbf{p}} - v_{\mathbf{k}} u_{\mathbf{p}}\right) \left(\alpha_{\mathbf{k},\uparrow}^+ \alpha_{-\mathbf{p},\downarrow}^+ + \alpha_{\mathbf{p},\uparrow} \alpha_{-\mathbf{k},\downarrow}\right).$$  (10)

The London penetration depth corresponds to the limit $q = 0$ [1,2]. In this case the energy operator (9) takes the form

$$H_{\text{kin}}(\mathbf{q}=0) = \sum_{\mathbf{k},\sigma} E_{\mathbf{k}} \alpha_{\mathbf{k},\sigma}^+ \alpha_{\mathbf{k},\sigma} + \frac{eA_{\mathbf{q}=0}^x}{\hbar c} \sum_{\mathbf{k}} \left(\frac{d\varepsilon_{\mathbf{k}}}{dk_x}\right) \left(\alpha_{\mathbf{k},\downarrow}^+ \alpha_{\mathbf{k},\downarrow} - \alpha_{-\mathbf{k},\uparrow}^+ \alpha_{-\mathbf{k},\uparrow}\right).$$  (11)

Hence, we find Bogoliubov's quasiparticle energies in the uniform vector-potential:

$$E_{\mathbf{k}}^{\downarrow} = E_{\mathbf{k}} + \frac{eA_{\mathbf{q}=0}^x}{\hbar c} \frac{d\varepsilon_{\mathbf{k}}}{dk_x},$$

$$E_{-\mathbf{k}}^{\uparrow} = E_{-\mathbf{k}} - \frac{eA_{\mathbf{q}=0}^x}{\hbar c} \frac{d\varepsilon_{\mathbf{k}}}{dk_x}.$$  (12)

The obtained equations are the natural generalization of the well-known equations as obtained in the weak coupling limit (see, e.g., equation (3.108) in [2]). Moreover, the form (12) presented by us is quite simple and useful from a physical point of view. In fact, it sheds new light on the fine detail of the interaction of Bogoliubov's quasiparticles with the vector-potential.

The mean value of the current is given by

$$j_x^p(\mathbf{q}=0) = \frac{e}{\hbar} \sum_{\mathbf{k}} \frac{d\varepsilon_{\mathbf{k}}}{dk_x} \left( \langle \alpha_{\mathbf{k},\downarrow}^+ \alpha_{\mathbf{k},\downarrow} \rangle - \langle \alpha_{-\mathbf{k},\uparrow}^+ \alpha_{-\mathbf{k},\uparrow} \rangle \right) = \frac{e}{\hbar} \sum_{\mathbf{k}} \frac{d\varepsilon_{\mathbf{k}}}{dk_x} \left[ f(E_{\mathbf{k}}^{\downarrow}) - f(E_{-\mathbf{k}}^{\uparrow}) \right].$$  (13)

The Fermi distribution functions $f(E_k^\sigma)$ can be expanded up to the linear terms,

$$f(E_{\mathbf{k}}^{\uparrow}) = \frac{1}{1+\exp\left(E_{\mathbf{k}}^{\uparrow}/k_B T\right)} \cong \frac{1}{1+\exp\left(E_{\mathbf{k}}/k_B T\right)} + \frac{df(E_{\mathbf{k}})}{dE_{\mathbf{k}}} \frac{eA_{\mathbf{q}=0}^x}{\hbar c} \frac{d\varepsilon_{\mathbf{k}}}{dk_x}.$$  (14)

Substituting (14) in (13) we obtain

$$j_x^p(\mathbf{q}=0) = -A_{\mathbf{q}=0}^x \frac{2e^2}{\hbar^2 c} \sum_{\mathbf{k}} \left(\frac{d\varepsilon_{\mathbf{k}}}{dk_x}\right)^2 \frac{df(E_{\mathbf{k}})}{dE_{\mathbf{k}}}.$$  (15)

In the weak coupling approximation equation (15) coincides with that given in [2]. The full equation for superfluid density given in [2], taking into account both the paramagnetic and diamagnetic currents, has the form (the second term in equation (3.111) in [2]):

$$\lambda_{\text{L}}^{-2}(T) = \lambda_{\text{L}}^{-2}(0)\left[1 - 2\int_{\Delta}^{\infty}\left(-\frac{\mathrm{d}f(E)}{\mathrm{d}E}\right)\frac{E}{(E^2 - \Delta^2)^{1/2}}\mathrm{d}E\right]. \tag{16}$$

Equation (15) should be compared with the second term in (16). It can be obtained from (13) only in the case where the values of the derivatives $(\mathrm{d}\varepsilon_{\mathbf{k}}/\mathrm{d}k_x)^2$ are equal at all points of the Fermi surface. In strong coupling superconductors this is not the case and, in particular, in copper oxide HTSC, this assumption is not true.

**4. Mean value of the diamagnetic current**

The derivation scheme is as follows. We write the correction to kinetic energy, which is quadratic over the vector-potential,

$$\delta H_{\text{kin}}^{(2)} = -\frac{1}{2}\left(\frac{e}{c\hbar}\right)^2 \sum_{n,l,\sigma} t_{nl}(A_x R_{nl}^x)^2 a_{n,\sigma}^+ a_{l,\sigma}. \tag{17}$$

Then we turn to Bogoliubov's quasiparticle operators and take the average over the ground state of a superconductor. Doing so for the component for the diamagnetic current we get:

$$\begin{aligned}
j_x^d(\mathbf{q}=0) &= -A_{\mathbf{q}=0}^x \frac{e^2}{\hbar^2 c}\sum_{\mathbf{k},\sigma}\frac{\mathrm{d}^2\varepsilon_{\mathbf{k}}}{\mathrm{d}(k_x)^2}\langle a_{\mathbf{k},\sigma}^+ a_{\mathbf{k},\sigma}\rangle \\
&= -A_{\mathbf{q}=0}^x \frac{e^2}{\hbar^2 c}\sum_{\mathbf{k}}\frac{\mathrm{d}^2\varepsilon_{\mathbf{k}}}{\mathrm{d}(k_x)^2}\left\langle \left(u_{\mathbf{k}}\alpha_{\mathbf{k},\uparrow}^+ - v_{\mathbf{k}}\alpha_{-\mathbf{k},\downarrow}\right)\left(u_{\mathbf{k}}\alpha_{\mathbf{k},\uparrow} - v_{\mathbf{k}}\alpha_{-\mathbf{k},\downarrow}^+\right) + \ldots\right\rangle \\
&= -2A_{\mathbf{q}=0}^x \frac{e^2}{\hbar^2 c}\sum_{\mathbf{k}}\frac{\mathrm{d}^2\varepsilon_{\mathbf{k}}}{\mathrm{d}(k_x)^2}\left(\frac{u_{\mathbf{k}}^2 - v_{\mathbf{k}}^2}{\exp(E_{\mathbf{k}}/k_{\text{B}}T)+1} + v_{\mathbf{k}}^2\right).
\end{aligned} \tag{18}$$

In the weak coupling limit the second derivative, $\mathrm{d}^2\varepsilon_{\mathbf{k}}/\mathrm{d}(k_x)^2$, is wavevector independent. In this case the sum $\sum_{\mathbf{k},\sigma}\langle a_{\mathbf{k},\sigma}^+ a_{\mathbf{k},\sigma}\rangle$ is the number of current carriers. It is temperature independent and corrensponds to the unity in the right hand side in equation (16). In the tight binding scheme the second derivative $\mathrm{d}^2\varepsilon_{\mathbf{k}}/\mathrm{d}(k_x)^2$ is not a fixed number and hence the diamagnetic current is temperature dependent. This is the second argument why a direct application of equation (15) is not suitable for the analysis of the temperature dependence of the London penetration depth in HTSC.

**5. The expression for superfluid density**

For numerical evaluations it is convenient to transform equation (18) as follows. The summation in (18) can be replaced by integration by introducing the density of states and taking the integral by parts afterwards. Taking into account the fact that the density of states at the top and at the bottom of the band is zero, we get the diamagnetic contribution:

$$j_x^d(\mathbf{q}=0) = 2\frac{e^2}{c\hbar^2}A_{\mathbf{q}=0}^x \sum_{\mathbf{k}}\frac{\mathrm{d}\varepsilon_{\mathbf{k}}}{\mathrm{d}k_x}\left\{\frac{\mathrm{d}}{\mathrm{d}k_x}\left[\frac{u_{\mathbf{k}}^2 - v_{\mathbf{k}}^2}{\exp(E_{\mathbf{k}}/k_{\text{B}}T)+1} + v_{\mathbf{k}}^2\right]\right\}. \tag{19}$$

Substituting $u_{\mathbf{k}}^2 = \tfrac{1}{2}\left[1 + (\varepsilon_{\mathbf{k}} - \mu)/E_{\mathbf{k}}\right]$ and $v_{\mathbf{k}}^2 = \tfrac{1}{2}\left[1 - (\varepsilon_{\mathbf{k}} - \mu)/E_{\mathbf{k}}\right]$ in (19) and combining the result with (15) one obtains the total current $j_x = j_x^p(\mathbf{q}=0) + j_x^d(\mathbf{q}=0)$, which can be compared with the London equation. Doing so we finally obtain the following expression:

$$\frac{1}{\lambda^2} = 4\pi\left(\frac{e}{c\hbar}\right)^2\left\{\sum_{\mathbf{k}}\frac{\mathrm{d}\varepsilon_{\mathbf{k}}}{\mathrm{d}k_x}\left[\frac{|\Delta_{\mathbf{k}}|^2}{E_{\mathbf{k}}^2}\frac{\mathrm{d}\varepsilon_{\mathbf{k}}}{\mathrm{d}k_x} - \frac{(\varepsilon_{\mathbf{k}} - \mu)}{2E_{\mathbf{k}}^2}\frac{\mathrm{d}|\Delta_{\mathbf{k}}|^2}{\mathrm{d}k_x}\right]\left[\frac{1}{E_{\mathbf{k}}} - \frac{\mathrm{d}}{\mathrm{d}E_{\mathbf{k}}}\right]\tanh\left(\frac{E_{\mathbf{k}}}{2k_{\text{B}}T}\right)\right\}. \tag{20}$$

It is in agreement with Refs. [4,7,10]. Note that equation (20) contains the modulus of the superconducting gap and, therefore, is independent of the phase of the order parameter $\Delta_{\mathbf{k}} = |\Delta_{\mathbf{k}}|e^{i\varphi_{\mathbf{k}}}$, as it should be in the gauge invariant theory [1]. It is clear also that at $T > T_{\text{c}}$ the quantity $1/\lambda^2$ (superfluid density) is zero, as it should be.

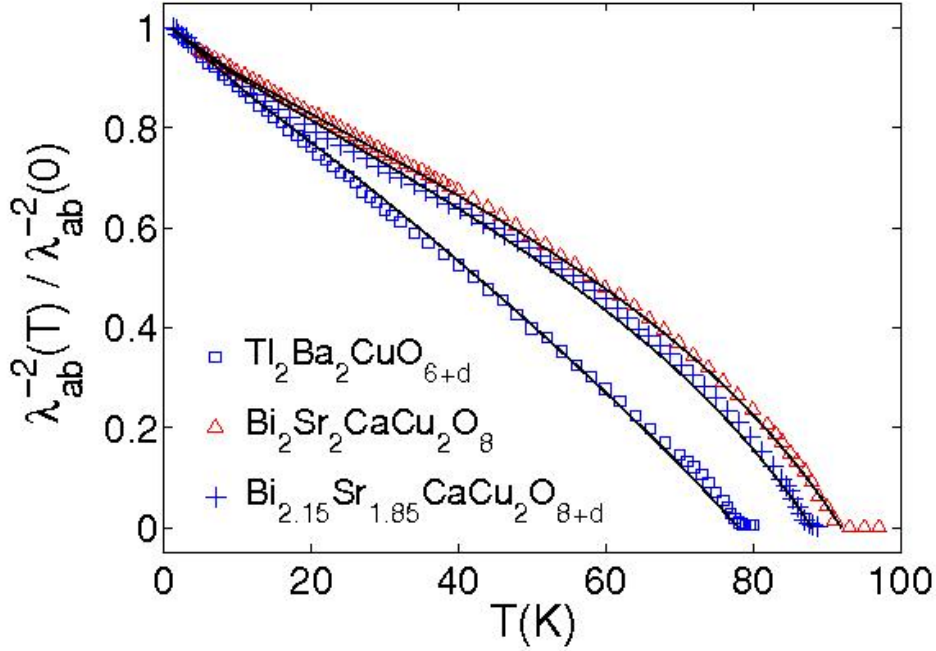

Figure 1. Temperature dependence of the superfluid density. Symbols: experimental data in single layer tetragonal compound $Tl_2Ba_2CuO_{6+\delta}$ ($T_c$=78 K) [29,40], in $Bi_{2.15}Sr_{1.85}CaCu_2O_{8+x}$ at optimal doping, $\delta$ = 0.16 (maximum $T_c$=87 K) [20] and in $Bi_2Sr_2CaCu_2O_8$ ($T_c$=93 K) [29,41]. Solid lines show the results of the calculations with $\Delta_d$=15 meV, $\Delta_d$=24 meV and $\Delta_d$=26 meV, respectively, and $\delta_t$ = 0. The energy dispersion parameters for $Tl_2Ba_2CuO_{6+\delta}$ are (in eV): $\mu$ = - 0.244, $t_1$= - 0.725, $t_2$= 0.302, $t_3$= 0.0159, $t_4$= -0.0805 and $t_5$ =0.0034 [42], and for both BiSrCaCuO samples extracted by Norman [21] (first hoppings set, in eV) are as follows: $\mu$ = - 0.1305, $t_1$= -0.5951, $t_2$= 0.1636, $t_3$ = -0.0519, $t_4$ = -0.1117, and $t_5$ = 0.0510.

## 6. Numerical results and discussion

We compare our calculations first with recent experimental data in BiSrCuO compounds [20]. Figure 1 shows the results of the calculations (solid lines, equation (20)). Symbols show the experimental data. We take the energy dispersion following [21], where the numerical values of the hopping integrals were defined from ARPES data. We take the temperature dependence of the superconducting gap parameter as extracted from the temperature dependence of the Cu and O Knight shift and the spin-lattice relaxation behaviour [22],

$$\Delta_\mathbf{k}(T) = \frac{\Delta_d}{2}(\cos k_x a - \cos k_y a)\tanh\left(\alpha\sqrt{\frac{T_c}{T}-1}\right), \qquad (21)$$

where $\Delta_d \cong 24$ meV and $\alpha \cong 1.76$. The characteristic feature of our theory is the linear behaviour at low temperatures. At this point one may treat the coincidence of the calculations and experimental data as a proof for d-wave pairing. We note that the analogous conclusion has been made for the first time in [23] from the $1/\lambda^2$ temperature dependence in $YBa_2Cu_3O_{6.95}$. However, it needs verification since the $1/\lambda^2$ analysis [23] used the equation with the effective mass approximation and has no connection with the actual energy dispersion in copper oxide HTSC.

We want to point out an important feature of HTSC compounds that the $2\Delta_m(T=0)/k_BT_c$ value has a strong effect on the $1/\lambda^2$ temperature dependence close to $T_c$. Here index "m" means a maximum value of the gap. The curvature of this dependence can be used for determination of the $2\Delta_m(T=0)/k_BT_c$ ratio in copper oxide HTSC compounds. Figure 2 illustrates this effect, showing the calculated $1/\lambda^2$ using equation (20) for a set of the $\Delta_d$ values. The $1/\lambda^2$ temperature dependence from [20] (figure 1) can be perfectly described by equation (20) with the energy dispersion defined from ARPES data [17] and $2\Delta_m(T=0)/k_BT_c$ ranges from 4.5 to 6.5.

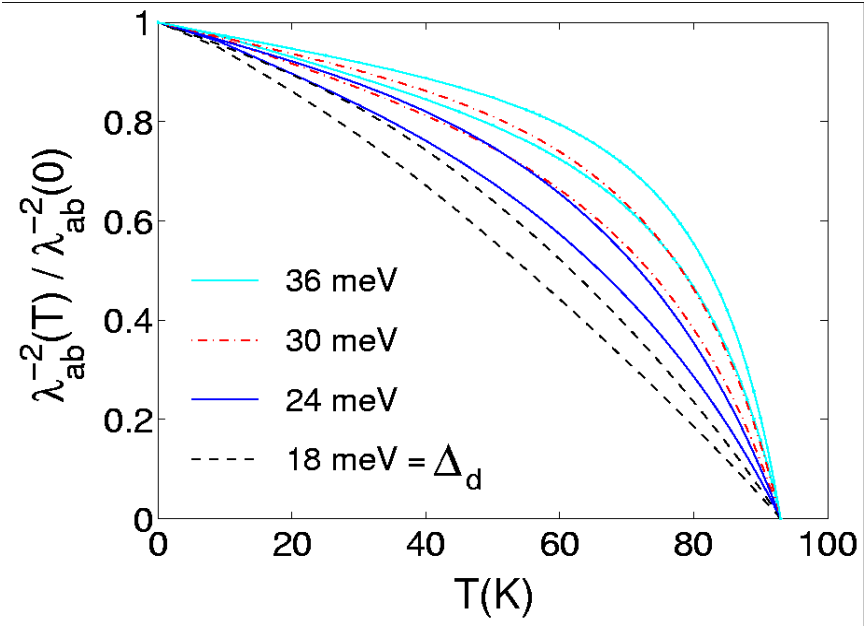

Figure 2. The calculated superfluid density $n_s \propto 1/\lambda^2$ *versus* temperature at various values of $\Delta_d$ = 18, 24, 30 and 36 meV, which corresponds to $2\Delta_d/k_B T_c$ = 4.5; 6; 7.5; 9, respectively, from down to up with $\alpha \cong 1.76$ and Norman second hopping parameter set [21], in eV: $\mu$ = - 0.1960, $t_1$ = -0.6798, $t_2$ = 0.2368, $t_3$ = -0.0794, $t_4$ = 0.0343, and $t_5$ = 0.0011. The lower curves of the same type show the tetragonal case. The neighbouring upper curves of the same type show the calculations with the same fixed $\Delta_d$ and in the orthorhombic case: $\delta_t$ = - 0.03, $\alpha' \cong \alpha$, $\Delta_s \cong 0.2\Delta_d$ and $\Delta_{ph} \cong 0.2\Delta_d$.

When comparing with experimental data it is important to note the following. The BiSrCuO and YBaCuO compounds are not tertagonal. The presence of orthorhombic distorsions leads to an admixture of s-wave component in the superconducting gap parameter. The analysis of the integral gap equation and the symmetry considerations lead to the conclusion that equation (21) should be replaced by the following form for the superconducting gap [19,24]:

$$\Delta_{\mathbf{k}}(T) = \left[\frac{\Delta_d}{2}(\cos k_x a - \cos k_y b) + \frac{\Delta_s}{2}(\cos k_x a + \cos k_y b)\right]\tanh\left[\alpha\sqrt{(T_c/T)-1}\right] + \Delta_{ph}(T). \quad (22)$$

The superconducting gap parameter becomes multicomponent. The $\Delta_{ph}$ component is, probably, due to the phonon mediated interaction. Its temperature dependence can be quite complicated. Below, for simplicity, we approximate it in the form $\Delta_{ph}(T) = \Delta_{ph}\tanh\left[\alpha'\sqrt{(T_c/T)-1}\right]$. From the semiempirical estimations based on the photoemission data [25], neutron scattering [19,24], tunnelling [26] and Raman [27] spectroscopies in YBaCuO family compounds, $\Delta_{ph} \cong 0.2\Delta_d$. From photoemission data [25], $\Delta(k_x a \cong \pi, k_y b = 0) \cong 28$ meV, $\Delta(k_x a \cong 0, k_y b = \pi) \cong 41$ meV, it follows $\Delta_s \cong 0.2\Delta_d$. The results of the calculations are shown in figures 2 and 3. The hopping integrals parameters are taken from [21]. The orthorhombicity parameter $|\delta_t| \cong 0.03$ (see [25,28]). As one can see, the presence of a small admixture of s-wave components in the superconducting gap parameter does not qualitatively affect the reduced temperature behaviour of the superfluid density $n_s \propto 1/\lambda^2$. However, it is clear that the effect of orthorhombicity is very essential for $\lambda_\alpha^{-2}(T)$ absolute values [29]. It would be informative for multicomponent superconductivity to study this effect experimentally.

One of the most outstanding properties of HTSC is the presence of the isotope effect in the magnetic field penetration depth parameter in a superconductor. As was emphasized in the pioneering paper [30] (the research review can be found in [31]), this effect gives important information about the interaction of a subsystem of charge carriers with phonons and an indication of the polaronic character of conductivity in these compounds. The ordinary superconductors do not possess this effect. Despite the evident importance of the isotope effect in $1/\lambda^2$, its detailed interpretation meets serious difficulties [32].

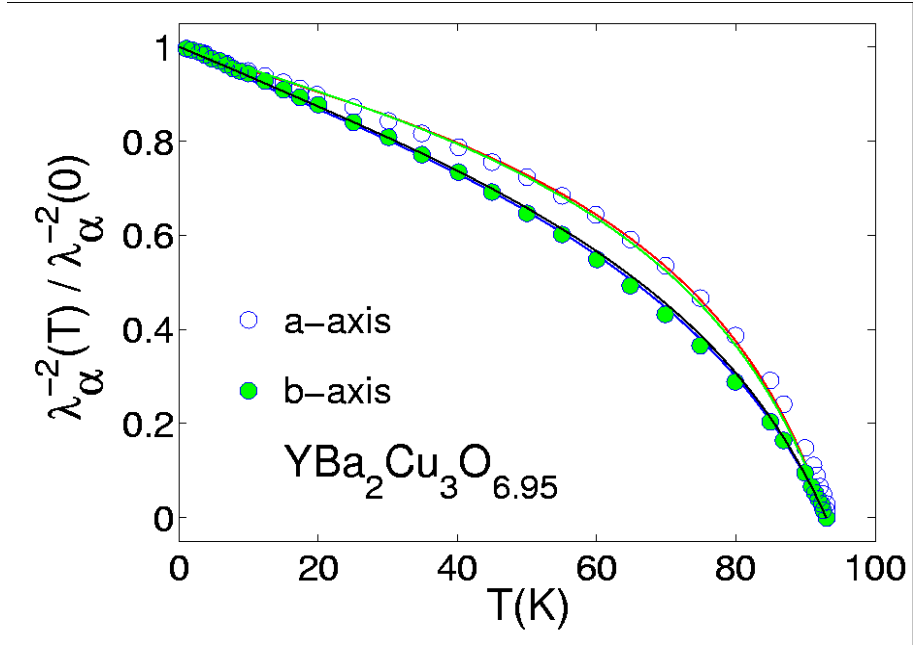

Figure 3. Temperature dependence of the superfluid density $n_s \propto 1/\lambda^2$ in optimally doped $YBa_2Cu_3O_{7-y}$ in the $a$ and $b$ directions. The experimental data is from Refs. [29,43]. Solid lines show the calculated $1/\lambda^2$ for both $\delta_t = -0.03$ and $\delta_t = 0.03$ and with the following parameters set: $\Delta_d = 25$ meV, $\alpha \cong 2$, the energy dispersion is taken as extracted by Norman [21] (second hoppings set, in eV): ($\mu = -0.1960$, $t_1 = -0.6798$, $t_2 = 0.2368$, $t_3 = -0.0794$, $t_4 = 0.0343$, and $t_5 = 0.0011$. The extracted relations are $\Delta_s \cong 0.1\,\Delta_d$ and $\Delta_{ph} \cong 0.1\,\Delta_d$. The analysis shows that one cannot distinguish between $\delta_t = -0.03$ and $\delta_t = 0.03$ from normalized $\lambda_\alpha^{-2}(T)/\lambda_\alpha^{-2}(0)$ behaviour.

Let us discuss the experimental data for the isotope effect in the $YBa_2Cu_3O_{7-\delta}$ superconductor. Figure 4 shows the temperature dependence of the superfluid density following equation (20) (solid lines). According to equation (20) one can separate the two reasons for the $^{16}O$ - $^{18}O$ isotope effect in the penetration depth. The first one is related to changes in the superconducting gap parameter $^{18}\Delta_m = {}^{16}\Delta_m (1 - \alpha_\Delta \Delta m/m)$, where $\alpha_\Delta$ is the experimentally measured parameter from $^{18}T_c - {}^{16}T_c = -\alpha_O\, {}^{16}T_c\, \Delta m/m$. Its origin is mainly related to $\Delta_{ph}$ component, which, in accord with the Bardeen-Cooper-Schrieffer (BCS) theory, is proportional to the Debye frequency. An additional source for the isotope effect is related to polaron renormalization of the superexchange coupling parameter [33]. According to the experimental data [34], the changes in $T_c$ values are small, $\alpha_O = 0.024(8)$, whereas the total isotope effect for the penetration depth, $\beta_O = -\dfrac{d\ln\lambda}{d\ln m}$, is $\beta_O(YBa_2Cu_3O_7) \cong -0.21(4)$. The experimental data can be fairly well explained if one assumes the change in the effective hopping parameters, $t$, by $^{16}O$ - $^{18}O$ exchange as $^{18}t = {}^{16}t(1 - \alpha_t \Delta m/m)$. Following the polaron theory [35,36] we suppose that the hopping's renormalization is independent from the distance between the sites. Accepting the above mentioned procedure as an algorithm for the determination of $\alpha_t$, we find $\alpha_t(YBa_2Cu_3O_7) = 0.35$. The same procedure using the experimental data for $La_{1.85}Sr_{0.15}CuO_4$ [37] gives $\alpha_t(La_{1.85}Sr_{0.15}CuO_4) = 0.26$.

It is instructive to compare the values of the coefficients for the conducting zone in $YBa_2Cu_3O_7$ and $La_{1.85}Sr_{0.15}CuO_4$ with the analogous parameters in $La_{0.75}Ca_{0.25}MnO_3$ and $Nd_{0.7}Sr_{0.3}MnO_3$ compounds. According to [38] in manganites the $^{16}O$ - $^{18}O$ isotope coefficients are $\alpha_{t^*}^O = 0.7$ and $\alpha_{t^*}^O = 1.1$, respectively. These values for manganites are 3-4 times larger compared with that extracted by us above for copper oxides. Qualitatively one may understand this as follows. The charge carriers move on the Mn sites. The polaronic band narrowing is caused mainly by shifts of the nearest oxygen ions. The oxygen mode is active. Since the charge carriers in hole doped HTSC are distributed over the oxygen positions the oxygen isotope effect is weak. The breathing mode of copper ions is active. In this case it will be instructive to perform the copper isotope effect on London penetration depth in hole doped copper oxide HTSC and the oxygen isotope effect in electron doped copper oxide PrCeCuO$_4$.

Finally, we want to note that our estimates for the hopping integral's renormalization due to the $^{16}$O - $^{18}$O isotope effect in YBa$_2$Cu$_3$O$_7$ does not contradict experimental data for the oxygen isotope effect on $T_c$. The isotope effect on hopping integrals leads to the renormalization of the Density of States (DoS) at the Fermi level and hence to a negative isotope effect on $T_c$ according to the BCS superconducting gap equation. The isotope effect on $T_c$ is usually positive, however, Ref. [39] reports the observation of the negative isotope effect on $T_c$. We note here that due to orthorhombic distortions the superconducting gap parameter gains an additional component, $\Delta_{ph}$, which gives a strong positive isotope effect on $T_c$, but relatively weakly affects $1/\lambda^2$. The polaronic renormalization of hoppings plays the dominant role in the isotope effect in $1/\lambda^2$. In this context either the smallness of the positive, or sometimes the observation of negative [39], isotope effects in $T_c$ becomes clearer. These effects are the consequences of two competing contributions. The contribution due to phonons gives a positive isotope $T_c$ shift, and polaronic narrowing of the conducting zone parameters gives a negative isotope effect on $T_c$. In this connection special interest arises for both $T_c$ and $1/\lambda^2$ isotope effect studies in copper oxide HTSC compounds without orthorhombic distortions, e.g., in Tl$_2$Ba$_2$CuO$_{6+\delta}$.

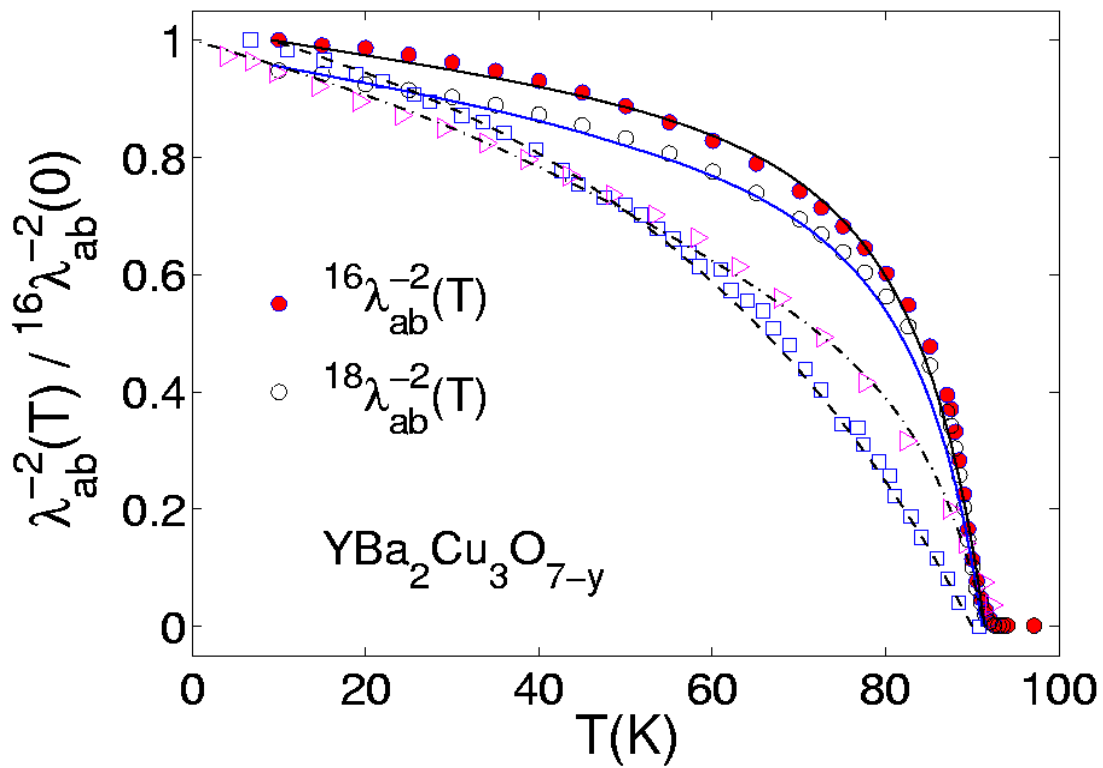

Figure 4. Temperature dependence of the superfluid density $n_s \propto 1/\lambda^2$ in optimally doped YBa$_2$Cu$_3$O$_{7-y}$. Circles show the isotope effect measured in [34]. Solid lines show the calculated $1/\lambda^2$ with the following parameters set: $\Delta_d$ = 40 meV, $\alpha \cong 1.76$, $\delta_t$ = - 0.03 and the energy dispersion is taken as extracted by Norman [21] (second hoppings set, in eV): $\mu$ = - 0.1960, $t_1$ = -0.6798, $t_2$ = 0.2368, $t_3$ = -0.0794, $t_4$ = 0.0343, $t_5$ = 0.0011. The extracted value for hoppings' $^{16}$O - $^{18}$O isotope renormalization is $\alpha_t$ = 0.35. The triangles and squares show the data from Refs. [23] and [44], respectively. The dashed line shows the results of the calculations, where $\Delta_d$ has been changed to $\Delta_d$ = 20 meV and the dash-dotted line with $\Delta_d$ = 20 meV, and with $\alpha \cong 2.9$ and $\alpha' \cong 2.5$. The relations $\Delta_s \cong 0.2\Delta_d$ and $\Delta_{ph} \cong 0.2\Delta_d$ are always fixed.

## 7. Conclusion

In conclusion, we present a simple derivation of an expression for superfluid density in the tight binding scheme, which, we hope, is understandable by a wide audience. We hope it will clarify some puzzles in the interpretation of experimental data in layered cuprates. Our analysis for temperature dependencies of the superfluid density $n_s \propto 1/\lambda^2$ shows that its curvature is very sensitive to the ratio $2\Delta_m(T=0)/k_B T_c$. The experimental data for $\lambda_{ab}^{-2}(T)$ in overdoped compound $Tl_2Ba_2CuO_{6+\delta}$ ($T_c$=78 K) fits fairly well with $2\Delta_m(T=0)/k_B T_c \cong 4.5$, whereas for optimally doped $Bi_2Sr_2CaCu_2O_8$ ($T_c$=93 K) the quantity $2\Delta_m(T=0)/k_B T_c \cong 6.5$. Different experimental methods for YBaCuO compounds near optimal doping level yield a quite different form for the temperature behaviour of $\lambda_{ab}^{-2}(T)$ (see figure 4). However, the fits of experimental data from Refs. [23] and [44] in fact give the same value: $2\Delta_m(T=0)/k_B T_c \cong 5.5$. The ratios extracted by us for $2\Delta_m(T=0)/k_B T_c$ are in agreement with findings for this quantity from experimental data. In particular, according to photoemission data [45] for optimally doped $Bi_2Sr_2CaCu_2O_8$ the value for this ratio is $6.1$, whereas the recent STM data [46] gives $2\Delta_m(T=0)/k_B T_c = 7.6$. Our calculated value 6.5 from the temperature dependence of penetration depth lies between these data.

The orthorhombic distortions affect the curvature of the temperature dependence of $\lambda_{ab}^{-2}(T)$. The $\lambda_\alpha^{-2}(T)$ anisotropy data in the a-b plane allows us to extract the admixtures of the anisotropic s-wave, $\Delta_s \approx 0.1\, \Delta_d$, and the isotropic s-wave, $\Delta_{ph} \approx 0.1\, \Delta_d$, components from the predominant d-wave component of the superconducting gap in $YBa_2Cu_3O_{6.95}$. The extracted value for hopping $^{16}O$ - $^{18}O$ isotope renormalization in copper oxide HTSC is relatively small, $\alpha_t = 0.35$, compared to that for manganites $La_{0.75}Ca_{0.25}MnO_3$ and $Nd_{0.7}Sr_{0.3}MnO_3$.

**Acknowledgments**

It is a pleasure to thank R. Khasanov, J. R. Cooper and C. Panagopoulos for sharing their $1/\lambda^2$ data in electronic form. We thank H. Keller, I. Eremin, J. Roos and O. A. Anikeenok for useful discussions. This work was supported in part by the Russian Foundation for Basic Research, Grant # 09-02-00777-a and Swiss National Science Foundation, Grant IZ73Z0_128242. I.A.L. was supported by the Dynasty Foundation.

**References**
[1] Schriffer J R 1964 *Theory of Superconductivity* (New York: Benjamin)
[2] Tinkham M 1996 *Introduction to Superconductivity* (New-York: McGraw-Hill)
[3] Won H and Maki K 1994 *Phys. Rev.* B **49** 1397-402
[4] Misawa S 1995 *Phys. Rev.* B **51** 11791-97
[5] Scalapino D J, White S R and Zhang S C 1992 *Phys. Rev. Lett.* **68** 2830-33
[6] Xiang T and Wheatley J M 1996 *Phys. Rev. Lett.* **77** 4632-35
[7] Benfatto L, Caprara S and Di Castro C 2000 *Eur. Phys. J.* B **17** 95-102
[8] Wang Q H, Han J H and Lee D H 2001 *Phys. Rev. Lett.* **87** 077004
[9] Micnas R and Tobijaszewska B 2002 *J. Phys.: Condens. Matter* **14** 9631-49
    Bussmann-Holder A, Micnas R and Bishop A R 2004 *Phil. Mag.* **21** 1257-64
[10] Sheehy D E, Davis T P and Franz M 2004 *Phys. Rev.* B **70** 054510
[11] Kim M S, Skinta J A and Lemberger T R 2002 *Phys. Rev.* B **66** 064511
[12] Khasanov R, Shengelaya A, Maisuradze A, La Mattina F, Bussman-Holder A, Keller H and Muller K A 2007 *Phys. Rev. Lett.* **98** 057007
[13] Khasanov R, Strassle S, Di Castro D, Masui T, Miyasaka S, Tajima S, Bussman-Holder A and Keller H 2007 *Phys. Rev. Lett.* **99** 237601
[14] Khasanov R, Shengelaya A, Maisuradze A, Di Castro D, Savic I M, Weyeneth S, Park M S, Jang D J, Lee S I and Keller H 2008 *Phys. Rev.* B **77** 184512
[15] Atkinson W A and Carbotte J P 1995 *Phys. Rev.* B **52** 10601-9
[16] Sharapov S G and Carbotte J P 2006 *Phys. Rev.* B **73** 094519
[17] Damascelli A, Hussain Z and Shen Z X 2003 *Rev. Mod. Phys.* **75** 473-541

[18] Feynman R, Leighton R and Sands M 1964 The Feynman Lectures on Physics vol 3 (London: Addison-Wesley)
[19] Shnyder A P, Manske D, Mudry C and Sigrist M 2006 *Phys. Rev.* B **73** 224523
[20] Anukool W, Barakat S, Panagopoulos C and Cooper J R 2009 *Phys. Rev.* B **80** 024516
[21] Norman M R 2007 *Phys. Rev.* B **75** 184514
[22] Mayer T, Eremin M, Eremin I and Meyer P F 2007 *J. Phys.: Condens. Matter* **19** 116209
[23] Hardy W N, Bonn D A, Morgan D C, Liang R and Zhang K 1993 *Phys. Rev. Lett.* **70** 3999-4002
[24] Eremin I and Manske D 2005 *Phys. Rev. Lett.* **94** 067006
[25] Lu D H *et al* 2001 *Phys. Rev. Lett.* **86** 4370-3
[26] Kirtley J R, Tsuei C C, Ariando A, Verwijs C J M, Harkema S and Hilgenkamp H 2006 *Nature Physics* **2** 190-4
[27] Bakr M, Schnyder A P, Klam L, Manske D, Lin C T, Keimer B, Cardona M and Ulrich C 2009 *Phys. Rev.* B **80** 064505
[28] Xiao-Shen Ye and Jian-Xin Li 2007 *Phys. Rev.* B **76** 174503
[29] Hardy W N, Kamal S and Bonn D A 1998 Magnetic penetration depths in cuprates: A short review of measurement techniques and results *The Gap Symmetry and Fluctuations in High Temperature Superconductors* (Cargese, Corsica, France, 1-13 September 1997) (NATO ASI Series B: Physics vol 371) ed J Bok *et al* (New York: Kluwer Academic / Plenum Publishers) pp 373-402
[30] Zhao G M, Hunt M B, Keller H and Muller K A 1997 *Nature* **385** 236-9
[31] Zhao G M, Conder K, Keller H and Muller K A 1998 *J. Phys.: Condens. Matter* **10** 9055-66
[32] Lyubin I E, Eremin M V, Eremin I M and Keller H 2009 Towards the theory of isotope effect of the London penetration depth in cuprates *Preprint* cond-mat/0902.1029
[33] Eremin M V, Eremin I M, Larionov I A and Terzi A V 2002 *JETP Lett.* **75** 395-8
[34] Khasanov R *et al* 2004 *Phys. Rev. Lett.* **92** 057602
[35] Kugel K I and Khomskii D I 1980 *Zh. Eksp. Teor. Fiz.* **79** 987 [1980 *Sov. Phys. JETP* **52** 501]
[36] Alexandrov A S and Mott N F 1995 *Polarons and Bipolarons* (Singapore: World Scientific)
[37] Khasanov R, Shengelaya A, Conder K, Morenzoni E, Savic I M, Karpinski J and Keller H 2006 *Phys. Rev.* B **74** 064504
[38] Alexandrov A S, Zhao G M, Keller H, Lorenz B, Wang Y S and Chu C W 2001 *Phys. Rev.* B **64** 140404
[39] Franck J P and Lawrie D D 1995 *J. Supercond.* **8** 591-4
[40] Broun D M, Morgan D C, Ormeno R J, Lee S F, Tyler A W, Mackenzie A P and Waldram J R 1997 *Phys. Rev.* B **56** R11443-R11446
[41] Lee S-F, Morgan D C, Ormeno R J, Broun D M, Doyle R A, Waldram J R and Kadowaki K 1996 *Phys. Rev. Lett.* **77** 735-8
[42] Plate M *et al* 2005 *Phys. Rev. Lett.* **95** 077001
[43] Zhang K, Bonn D A, Kamal S, Liang R, Baar D J, Hardy W N, Basov D and Timusk T 1994 *Phys. Rev. Lett.* **73** 2484-7
[44] Stajic J, Iyengar A, Levin K, Boyce B R and Lemberger T R 2003 *Phys. Rev.* B **68** 024520
[45] Gu C *et al* 1995 *Phys. Rev.* B **51** 1397-1400
[46] Pushp A, Parker C V, Pasupathy A N, Gomes K K, Ono Sh, Wen J, Xu Zh, Gu G, Yazdani A 2009 *Science* **324** 1689